\documentclass[conference]{IEEEtran}
\usepackage[
top    = 0.7in,
bottom = 1in,
left   = 1.69cm,
right  = 1.69cm]{geometry}
\ifCLASSINFOpdf
\else
\fi
\usepackage{graphicx}
\usepackage{setspace}
\usepackage{amsmath}
\usepackage{amssymb} 
\usepackage{bm} 
\usepackage{cite}
\usepackage{float}
\usepackage[usenames,dvipsnames]{pstricks}
\usepackage{epsfig}
\usepackage{pst-grad} 
\usepackage{pst-plot} 
\usepackage{tabularx}
\usepackage{balance}
\usepackage[T1]{fontenc}

\allowdisplaybreaks
\allowdisplaybreaks
\usepackage{caption}
\usepackage{subcaption}
\usepackage[utf8]{inputenc}

 \IEEEaftertitletext{\vspace{-2\baselineskip}}
\begin{document}
\onecolumn{
\noindent © 2024 IEEE. Personal use of this material is permitted. Permission from IEEE must be obtained for all other uses, in any current or future media, including reprinting/republishing this material for advertising or promotional purposes, creating new collective works, for resale or redistribution to servers or lists, or reuse of any copyrighted component of this work in other works.
}
 \twocolumn{
\bstctlcite{IEEEexample:BSTcontrol}
\title{Multi-task
Learning-based Joint CSI Prediction and Predictive Transmitter Selection for Security 
     \vspace{-.3cm} }

\author{
    \IEEEauthorblockN{Shashi Bhushan Kotwal\IEEEauthorrefmark{1,2}, 
       Chinmoy Kundu\IEEEauthorrefmark{2} 
     Sudhakar Modem\IEEEauthorrefmark{3},
    Holger Claussen\IEEEauthorrefmark{4}, 
    and Lester Ho\IEEEauthorrefmark{5}
    }

    \IEEEauthorblockA{\IEEEauthorrefmark{1}\IEEEauthorrefmark{3} Department of EE, Indian Institute of Technology Jammu, Jammu \& Kashmir, India }
      \IEEEauthorblockA{\IEEEauthorrefmark{1}School of E\&CE, Shri Mata Vaishno Devi University, Katra, Jammu \& Kashmir, India}

    
     \IEEEauthorblockA{\IEEEauthorrefmark{2}\IEEEauthorrefmark{4}Wireless Communications Laboratory, Tyndall National Institute, Dublin, Ireland}
     \IEEEauthorblockA{\IEEEauthorrefmark{4}School of Computer Science and Information Technology, University College Cork, Cork, Ireland}

      \IEEEauthorblockA{\IEEEauthorrefmark{4}Trinity College Dublin, Dublin, Ireland}

    \textrm{
    {
     {\{\IEEEauthorrefmark{1}shashi.kotwal,
     \IEEEauthorrefmark{3}sudhakar.modem\}}@iitjammu.ac.in},  
     {\{\IEEEauthorrefmark{2}chinmoy.kundu,
     \IEEEauthorrefmark{4}holger.claussen,
     \IEEEauthorrefmark{5}lester.ho\}}@tyndall.ie    
     }
        \vspace{-.5cm} 
     
     }
\maketitle

\thispagestyle{plain}
 \pagestyle{plain}

\begin{abstract}
In mobile communication scenarios, the acquired channel state information (CSI) rapidly becomes outdated due to fast-changing channels. Opportunistic transmitter selection based on current CSI for secrecy improvement may be outdated during actual transmission, negating the diversity benefit of transmitter selection. Motivated by this problem, we propose a joint CSI prediction and predictive selection of the optimal transmitter strategy based on historical CSI by exploiting the temporal correlation among CSIs. The proposed solution utilizes the multi-task learning (MTL) framework by employing a single Long Short-Term Memory (LSTM) network architecture that simultaneously learns two tasks of predicting the CSI and selecting the optimal transmitter in parallel instead of learning these tasks sequentially. 
The proposed LSTM architecture outperforms convolutional neural network (CNN) based architecture due to its superior ability to capture temporal features in the data.  
Compared to the sequential task learning models, the MTL architecture provides superior predicted secrecy performance for a large variation in the number of transmitters and the speed of mobile nodes. 
It also offers significant computational and memory efficiency leading to a substantial saving in computational time by around 40 percent.  
\end{abstract}


\vspace{-0.2cm}
\section{Introduction}

Transmitter selection (TS) is an effective way to achieve diversity gain in a multi-transmitter system. Based on available channel state information (CSI), various transmitter selection schemes are implemented to utilize this diversity gain to improve the secrecy of communication networks \cite{Chinmoy_GC16, Chinmoy_letters21, Shalini_GC20, SBK_VTC21, sbk_tvt_2024}.
However, these TS techniques are only applicable in slow-fading channels.  In a practical mobile communication scenario, such as vehicle-to-everything, high-speed train, unmanned aerial vehicles, etc., the acquired CSI rapidly becomes obsolete due to the Doppler effect \cite{pred_chan_tas_vtc_2006}.

In a rapidly varying channel, the actual CSI during transmission may differ from its historical value applied to select a transmitter.  Transmission by a wrongly selected transmitter due to outdated CSI can degrade the secrecy performance of a system employing TS \cite{pred_chan_tas_vtc_2006,chan_pred_wcnc_21}. To alleviate this problem, transmitter selection is proposed based on predicted CSI in \cite{pred_chan_tas_vtc_2006}. 
The CSI is predicted using a traditional linear autoregressive (AR) stochastic model. The results show that transmitter selection based on predicted channels significantly improves the bit error compared to selection using outdated CSI.
The accuracy of the traditional AR approaches is bounded by the amount of historical CSI data and the computational complexity is proportional to the square of AR model order \cite{ ml_chan_pred_twt_20}.

 In contrast to the traditional AR models, machine learning (ML) models can provide more accurate CSI prediction by exploiting the inherent characteristics and hidden relationships within historical CSI data via non-linear approaches \cite{ml_chan_pred_twt_20, chan_pred_wcnc_21, chan_pred_mimo_imperfect_csi_icc_20, chan_pred_lstm_wcnc_22, dnn_tvt_2022}.  Using a Long Short-Term Memory (LSTM)-based deep learning ML model, highly accurate CSI is predicted which is subsequently used for relay selection in case of outdated CSI in \cite{chan_pred_wcnc_21}. 




CSI prediction and predictive transmitter selection (transmitter or relay) are different tasks. 
The task of transmitter selection in \cite{pred_chan_tas_vtc_2006} or the predictive relay selection in \cite{chan_pred_wcnc_21} 
rely on a two-step sequential approach 
(CSI prediction task followed by transmitter selection task). In contrast, in \cite{pred_tas_elsevier_24} a single-step approach is proposed using a convolutional neural network (CNN) to predict selected transmitter directly from historical CSI values. Directly predicting selected transmitter from the historical CSI can avoid two-step computation, however, it does not solve the problem of CSI prediction. CSI prediction is also required for supporting various link-adaptive schemes, such as adaptive modulation and coding, power adaptation, etc. and performance prediction. Additionally, CSI prediction can reduce the need for channel estimation, pilot transmissions, and feedback overhead, thus improving the channel usage efficiency.

The articles \cite{pred_chan_tas_vtc_2006,chan_pred_wcnc_21,pred_tas_elsevier_24} have focused on a single task at a time. However,  simultaneously performing both CSI prediction and predictive transmitter selection 
provide a suitable solution for meeting mobile networks' computational and memory constraints. It can reduce the training data requirement compared to that required when CSI prediction and predictive transmitter selection tasks are implemented in sequence.
The multi-task learning (MTL) framework can offer the solution for simultaneously learning multiple tasks in ML \cite{mtl_thesis_basic_97,ieee_conf_mtl_18,mtl_seq2seq_tvt_20}. MTL framework leverages the inherent relation among training data across different tasks, allowing the model to exploit these relations and potentially improve the performance of all tasks. 
By sharing the common parameters across different tasks, the training model complexity can also be reduced in MTL.

Motivated by the above discussion, a system with multiple transmitters and eavesdroppers in a mobile communication scenario is considered for joint CSI prediction and predictive transmitter selection for secrecy enhancement. 
It is to be noted that articles \cite{pred_chan_tas_vtc_2006,chan_pred_wcnc_21,pred_tas_elsevier_24} though considered predictive TS under outdated CSI condition, these articles did not consider secrecy. Moreover, none of these articles implemented multiple tasks jointly. Although \cite{chan_pred_mimo_imperfect_csi_icc_20} considered secrecy, the work was restricted to the CSI prediction task only.  The main contributions of our work are outlined below.
\begin{itemize}
    \item The joint CSI prediction and predictive transmitter selection problem is solved using MTL-based ML architecture that exploits the temporal relations in the CSI using LSTM model.

    \item We show that while both the LSTM and CNN architectures can be employed for predicting the CSI and the optimal transmitter jointly, LSTM outperforms CNN due to its superior ability to capture temporal features in the data.

    \item We show that the MTL-based model has better predicted secrecy over a large relative node speed and number of transmitters as compared to the sequential task learning model.
    

    \item In contrast to the sequential learning model, the MTL-based joint CSI prediction and transmitter selection architecture is computationally and memory efficient. It can save computation time by around 40 percent. 


\end{itemize}

The rest of the paper is organized as follows. The system model is described in Section \ref{section_System_Model}. The LSTM model for joint CSI prediction and predictive transmitter selection is described in Section \ref{sec_ml}, results are presented in Section \ref{section_results} and conclusions are provided in Section \ref{section_conclusion}.

\textit{Notation:}  $|\cdot|$ and $||\cdot||_2$ denote magnitude and $L_2$ norm operators, respectively.  $\mathbb{E}\left[\cdot \right]$ and 
 $R_{hh}(\cdot)$ denote the expectation and autocorrelation operator, respectively. $\mathbf{T}$ denotes the transpose operator and $\left[x\right]^+=\max\{0,x\}$ where $\max\{0,x\}$ denotes maximum of $0$ and $x$.           

 \section{System Model}\label{section_System_Model}\begin{figure}
 \centering
\includegraphics[width=4cm,height=4cm,keepaspectratio]{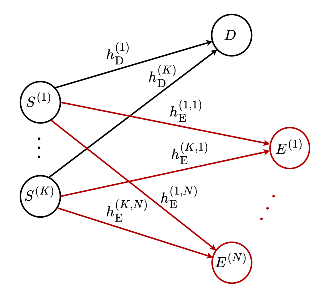}
 \vspace{-0.2cm}
 \caption{A system with $K$ transmitters and $N$  eavesdroppers for joint CSI prediction and predictive transmitter selcetion.}
 \label{system_model}
 \vspace{-0.5cm}
 \end{figure}
We consider a mobile wiretap communication system as in Fig. \ref{system_model} consisting of $K$ number of transmitters $\mathrm{S}^{(k)}$, where $k \in \{1, \ldots, K\}$, a legitimate destination D, and $N$ passive eavesdroppers $\mathrm{E}^{(n)}$, where $n \in \{1, \ldots, N\}$. 
It is assumed that all nodes are equipped with a single antenna. The $k$-th channel between $\mathrm{S}^{(k)}$ and $\mathrm{D}$ for each $k \in \{1, \ldots, K\}$ is assumed to undergo an independent and identically distributed (i.i.d.) fading. 
Due to the Doppler effect in the mobile scenario, the destination and eavesdropping channels are time-varying.
In this paper, the time-varying channels are modelled based on Clarke's model \cite{clarke_rayleigh_jour_68,time_corr_chan_data_sim_twc_2005}. Thus, the impulse response of the  time-varying channel $\mathrm{S}^{(k)}$-$\mathrm{D}$ can be stated with $M$ number of propagation paths as 
\begin{align}
\label{eq_sd_chan_model}
    {h}_{\mathrm{D}}^{(k)}(t)&=\frac{1}{\sqrt{M}}\left(Y_{CD}^{(k)}(t)+jY_{SD}^{(k)}(t)\right),
\end{align}
where $Y_{CD}^{(k)}(t)$ and $Y_{SD}^{(k)}(t)$ are the quadrature components and are expressed as
\begin{align}
\label{eq_qudrature1}
 Y_{CD}^{(k)}(t)&=\sum_{m=1}^{M}\cos\left(2 \pi f_d t \cos \beta_{m}^{(k)} + \phi_{m}^{(k)}\right),   \\
\label{eq_qudrature2}
 Y_{SD}^{(k)}(t)&=\sum_{m=1}^{M}\sin\left(2 \pi f_d t \cos \beta_{m}^{(k)} + \phi_{m}^{(k)}\right),
\end{align}
$f_d={f_c v}/{c}$ is the maximum Doppler frequency,  where $f_c$ is the carrier frequency, $v$ is the relative velocity between the transmitters and the receivers including legitimate destination and eavesdroppers, and $c$ is the speed of the light; and $\beta_{m}^{(k)}=({2 \pi m +\theta_{m}^{(k)}})/{M}$ and $\phi_{m}^{(k)}$ are the angle of arrival and initial phase of the $m$-th propagation path, respectively, at the destination. 
Both $\theta_{m}^{(k)}$ and $\phi_{m}^{(k)}$ are statistically independent and uniformly distributed over $\left[-\pi,\pi\right]$ for each $m$.  The autocorrelation function
of the channel complex envelope obtained from (\ref{eq_sd_chan_model}) is 
    $R_{hh}(\tau)=J_0(2 \pi f_d \tau)$, 
where $J_0(\cdot)$ is the zero-order Bessel function of the first kind \cite{time_corr_chan_data_sim_twc_2005}. The correlation in time between CSI samples is introduced due to the nature of the time functions defined in (\ref{eq_qudrature1}) and (\ref{eq_qudrature2}).
The coherence time $T_c$ of all the channels is $T_c={0.423}/{f_d}$ \cite{sklar_mag_97}.

The instantaneous SNR of the channel $\mathrm{S}^{(k)}$-$\mathrm{D}$ is given by $\gamma_{\mathrm{D}}^{(k)}(t)=\bar{\gamma}_{\mathrm{D}}|h_\mathrm{D}^{(k)}(t)|^2$ where $\mathbb{E}[|h_\mathrm{D}^{(k)}(t)|^2]=1$, $\Bar{\gamma}_\mathrm{D}=\alpha_\mathrm{D}  P_T/{(\sigma_\mathrm{D})}^2$ is the average received SNR at D, $P_T$ is the transmitted power, $\alpha_\mathrm{D}$ is the path loss factor inversely proportional to the distance between $\mathrm{S}^{(k)}$ and D, and $({\sigma_\mathrm{D}})^2$ is the average noise power at D.   Similar to ${h}_{\mathrm{D}}^{(k)}(t)$, the time-varying channel between $\mathrm{S}^{(k)}$ and $\mathrm{E}^{(n)}$, $h_\mathrm{E}^{(k,n)}(t)$, for each $k \in \{1, \ldots, K\}$ and $n \in \{1, \ldots, N\}$ is assumed to undergo an i.i.d. fading. The impulse response of the channel $\mathrm{S}^{(k)}$-$\mathrm{E}^{(n)}$ with $L$ number of propagation paths can also be modelled similar to ${h}_{\mathrm{D}}^{(k)}(t)$ as shown in (\ref{eq_sd_chan_model}). With $\mathbb{E}[|{h}_\mathrm{E}^{(k,n)}(t)|^2]=1$, the average SNR at $\mathrm{E}^{(n)}$ is expressed as $\Bar{\gamma}_\mathrm{E}=\alpha_\mathrm{E}  P_T/({\sigma_\mathrm{E}})^2$ where $\alpha_\mathrm{E}$ is the path loss factor inversely proportional to the distance between $\mathrm{S}^{(k)}$ and $\mathrm{E}^{(n)}$, and $({\sigma_\mathrm{E}})^2$ is the average noise power at $\mathrm{E}^{(n)}$.

To consider the worst possible eavesdropping scenario, we assume that the eavesdroppers are colluding and perform maximal ratio combining (MRC) of the received signals to extract confidential information.
The combined instantaneous SNR $\gamma_{\mathrm{E}}^{(k)}(t)$ at the eavesdroppers corresponding to the $k$-th transmitter is therefore given by
\begin{align}\label{eq_SNR_E_MRC}    \gamma_{\mathrm{E}}^{(k)}(t)&=\bar{\gamma}_E\sum_{n=1}^{N}|h_{\mathrm{E}}^{(k,n)}(t)|^2,
\end{align}
The optimal transmitter that provides the maximum instantaneous secrecy rate among all the transmitters is selected. The instantaneous secrecy rate of the system with optimal transmitter selection is defined in bits per channel use (bpcu) as \cite{SBK_VTC21}
\begin{align}\label{eq_sec_rate}  
C_{S}^{(k^*)}(t)&=\max_{k \in {1, \ldots, K}}\left[\log_2\left(\frac{1+\gamma_{\mathrm{D}}^{(k)}(t)}{1+\gamma_{\mathrm{E}}^{(k)}(t)}\right) \right]^+,
\end{align}
where $k^*$ is the optimal selected transmitter. The ergodic secrecy rate (ESR) of the system for the optimal transmitter selection is then given as \cite{sbk_tvt_2024}
\begin{align}
    \label{eq_ESR_CDF_eqn}
 C_{\mathrm{erg}}&= \mathbb{E}\left\{C_{S}^{(k^*)}(t)\right\}. 
\end{align}






\color{black}
\section{Joint  CSI Prediction and Predictive Transmitter Selection}
\label{sec_ml}
In this section, we formulate the problem of joint  CSI prediction and predictive transmitter selection. We then propose an LSTM-based MTL algorithm for the solution.

\subsection{Problem Formulation}
The joint CSI prediction and predictive transmitter selection is formulated as an MTL problem where CSI for future time instances and optimal transmitter that maximizes the instantaneous secrecy rate for those time instances are jointly predicted through a single ML network based on historical CSI.
To formally define the problem statement, we first define the CSI vector  that is used to propose the ML algorithm as
\begin{align}\label{eq_h_t}  
 \mathbf{h}(t) &=\Big[ \mathbf{h}_\mathrm{D}(t) , \mathbf{h}_\mathrm{E}(t) \Big]^\mathbf{T},
\end{align}
where
\begin{align}\label{eq_h_D_t}
 \mathbf{h}_\mathrm{D}(t) &=
\Big[ h_\mathrm{D}^{(1)}(t) ,\ldots, h_\mathrm{D}^{(K)}(t) 
\Big],\\
 \mathbf{h}_\mathrm{E}(t)  &=
\Big[ h_\mathrm{E}^{(1,1)}(t), \ldots,   h_\mathrm{E}^{(1,N)}(t), h_\mathrm{E}^{(2,N)}(t),\ldots,   h_\mathrm{E}^{(K,N)}(t)\Big],
\end{align}
and $|\mathbf{h}(t)|=\Big[ |\mathbf{h}_\mathrm{D}(t)|, |\mathbf{h}_\mathrm{E}(t)| \Big]^\mathbf{T}$ is the vector of length $K+(K\times N)$ with magnitudes of current CSI samples at time $t$. The proposed problem can then be formally stated as
\begin{align}\label{eq_prob_statement}
&\{|\mathbf{\hat{h}}(t+1)|,\hat{k}_1^*\},\ldots,\{|\mathbf{\hat{h}}(t+J)|,\hat{k}_J^*\}\nonumber\\ 
&=f\left(|\mathbf{h}(t-T+1)|, \ldots,|\mathbf{h}(t)|\right), \end{align}
where $T$ is the total length of the past CSI samples, $J$ is the length of the predicted CSI samples, $f(\cdot)$ is an arbitrary function that maps the $T$ historical CSI samples  to the future estimates of the channels $|\mathbf{h}(t+1)|,\ldots,|\mathbf{h}(t+J)|$ as $|\mathbf{\hat{h}}(t+1)|,\ldots,|\mathbf{\hat{h}}(t+J)|$ and also jointly predicts $J$ future selected transmitter indexes $k_1^*,\ldots,k_J^*$ as $\hat{k}_1^*,\ldots, \hat{k}_J^*$. The problem is then to design an ML model that implements the function $f(\cdot)$ in (\ref{eq_prob_statement}). The solution is provided using the LSTM-based MTL algorithm.


\subsection{Model Training Methodology and Multi-task loss function}

The prediction of the CSI based on historical CSI is implemented as a time series prediction problem \cite{pred_tc_22}. We utilize the temporal dependency capturing feature of LSTM to predict the CSI of the destination and eavesdroppers' channels.  Additionally, we use an LSTM-based time series classifier for predicting the selected transmitter.  For both time series prediction and time series classification, our proposed LSTM-based model exploits the temporal correlation in the wireless channel.

The proposed model, named as LSTM-J model, as shown in Fig. \ref{fig_ml_network_combined}, utilizes the MTL framework where a single LSTM layer is used followed by two Fully Connected Layers each serving different tasks \cite{mtl_thesis_basic_97}. The Fully Connected Layer-I assists in  CSI prediction whereas, the Fully Connected Layer-II assists in transmitter selection by providing its output to a softmax layer that converts the output into the probability distribution over the classes of the transmitters for transmitter classification.



We provide past CSI data as the input to the network as shown in (\ref{eq_prob_statement}) to train the model depicted in Fig. \ref{fig_ml_network_combined}. The input data for the training is arranged in $N_{data}$ training sets where each set consists of three matrices  $\mathbf{H_{train}}$, $\mathbf{H_{target}}$, and $\mathbf{K_{target}}$.  The matrix $\mathbf{H_{train}}^{K(1+N) \times T}=\left[|\mathbf{h}(t-T+1)|, \ldots,|\mathbf{h}(t)|\right]$ corresponds to the historical input data of destination and eavesdropping channel,  the matrix $\mathbf{H_{target}}^{K(1+N) \times J}=\left[|\mathbf{h}(t+1)|, \ldots,|\mathbf{h}(t+J)|\right]$ corresponds to the target future CSI values that follow the historical values $\mathbf{H_{train}}$, and the vector $\mathbf{K_{target}}^{1\times J}$ corresponds to the transmitters selected for $J$ future time instances according to the target future CSI values $\mathbf{H_{target}}^{K(1+N) \times J}$. The training is completed in mini batches where each batch contains $N_b$ training sets.


To design the multi-task loss function $L_{mt}$ for training, we choose individual loss functions, mean square error (MSE) loss function $L_p$ corresponding to the CSI prediction task and cross-entropy loss function $L_c$ corresponding to the transmitter classification task and produce a weighted sum of these two loss functions as MTL loss function. This is mathematically defined as  \cite{ieee_conf_mtl_18}
\begin{align}\label{eq_tot_loss}    
L_{mt}=  w L_p+ (1-w)L_c,
\end{align}
where $0<w<1$ is the weight assigned to $L_p$ and 
\begin{align}\label{eq_mse}    
L_p=\frac{1}{JK(N+1)}\sum_{n_{b}=1}^{N_{b}}\sum_{j=1}^{J}||\mathbf{e}_{n_b}^{(j)}||_2^2,\\
\label{eq_cross_entropy}   L_c=-\sum_{n_{b}=1}^{N_{b}}\sum_{k=1}^{K}p_{n_{b},k}\ln\left(q_{n_{b},k}\right).
\end{align} 
In (\ref{eq_mse}), $\mathbf{e}_{n_b}^{(j)}=(|\mathbf{\hat{h}}(t+j)|_{n_b}-|\mathbf{h}(t+j)|_{n_b})$ is the element-wise difference in the predicted CSI $|\mathbf{\hat{h}}(t+j)|_{n_b}$ and target CSI $|\mathbf{h}(t+j)|_{n_b}$ for the $j$-th time instance in the $n_{b}$-th training set where $j \in \{1,\ldots, J\}$ and $n_{b}\in\{1, \ldots, N_b\}$.
In (\ref{eq_cross_entropy}), $p_{n_{b},k}$ as well as $q_{n_{b},k}$ represent the predicted probability and actual probability, respectively, assigned to the $k$-th transmitter in the $n_{b}$-th training set for the transmitter classification purposes. The objective of the training is to minimize $L_{mt}$.
 



\color{red}



\color{black}

\subsection{Comparison with other MTL and sequential ML models}

We compare the performance of the proposed LSTM architecture with that of the CNN-based MTL architecture. In the CNN architecture, the LSTM layer in Fig. \ref{fig_ml_network_combined} is replaced by the combination of a one-dimensional CNN layer, a Rectified Linear Unit (ReLU) layer, a Layer Normalization layer, and a Global Average Pooling layer \cite{1D_cnn_classifier_tvt_22}.  In the CNN architecture, one-dimensional convolution filters stride across time to extract temporal features from the available CSI, enabling the prediction of future CSI and transmitter selection. The incorporation of non-linearity due to the ReLU layer in CNN architecture enables the network to learn complex representations of the data. We also compare the performance of LSTM-J model in Fig. \ref{fig_ml_network_combined} with that of the model in Fig. \ref{fig_ml_network_separate}, denoted as LSTM-S model, where two individual networks, one for CSI prediction task and the another for transmitter selection task based on predicted CSI, are employed sequentially.
\begin{figure}
 \centering    \includegraphics[width=5.5cm]{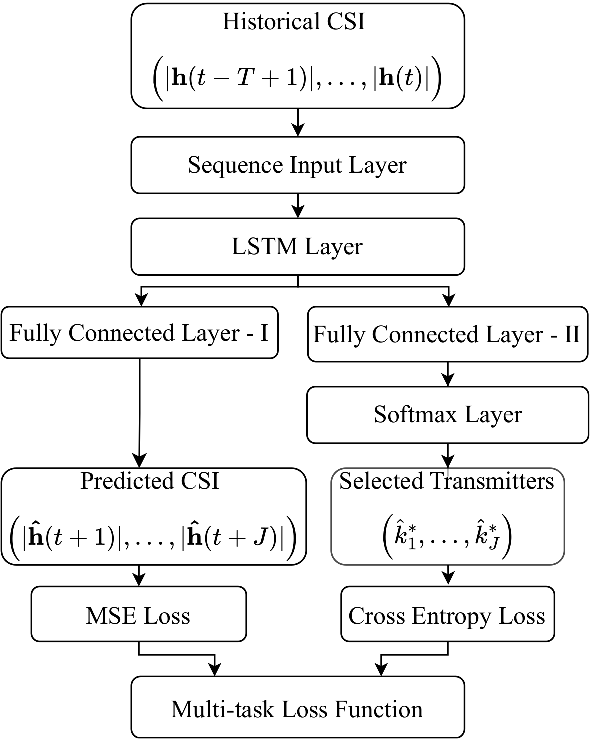}
 \caption{Layered architecture of the LSTM-J model trained for predicting both CSI and selected transmitter jointly.}  \label{fig_ml_network_combined}
 \vspace{-0.4cm}
 \end{figure} 
 \begin{figure}
 \centering \includegraphics[width=8.5cm]{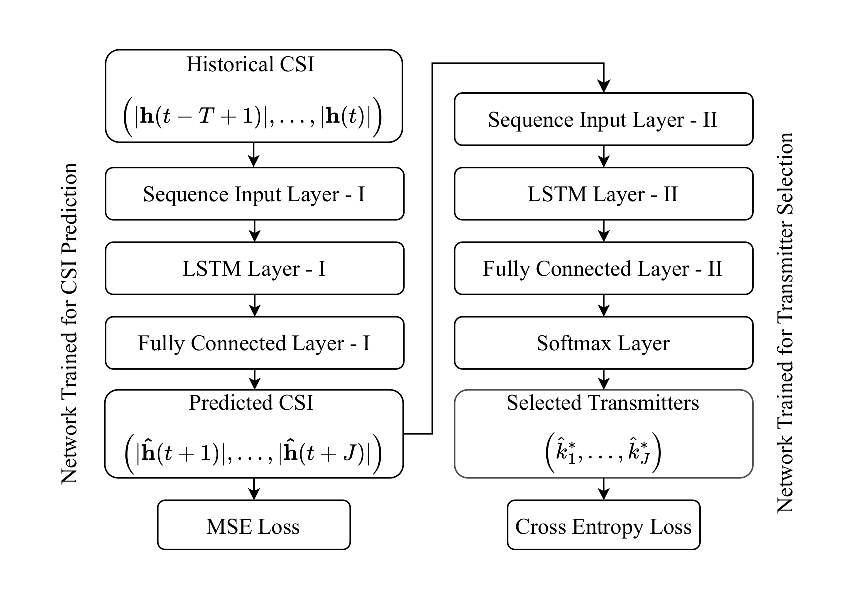}
  \vspace{-0.4cm} 
 \caption{Layered architecture of the LSTM-S model for predicting CSI and selected transmitter sequentially.}
 \vspace{-0.3cm} 
\label{fig_ml_network_separate}
 \end{figure}  
\subsection{Performance Metrics}
We evaluate the performance of proposed ML models for their CSI prediction accuracy and the accuracy in transmitter selection.  The CSI prediction accuracy is measured using normalised mean square error (NMSE) between future estimates of the channels and the actual channels, expressed as \cite{pimrc_19_seq_2_seq}
\begin{align}
\label{eq_NMSE}
    \textrm{NMSE}
    &=\frac{\mathbb{E}\{\sum_{j=1}^{J}\left(|||\mathbf{\hat{h}}(t+j)|-|\mathbf{h}(t+j)|||_2^2\right)\}}{\mathbb{E}\{\sum_{j=1}^{J}||\mathbf{h}(t+j)||_2^2\}}.
\end{align}
The transmitter selection accuracy is evaluated as the ratio of the number of instances when $k^*=\hat{k}^*$ to the total number of instances, expressed as a percentage.

Using $\{|\mathbf{\hat{h}}(t+1)|,\hat{k}_1^*\},\ldots,\{|\mathbf{\hat{h}}(t+J)|,\hat{k}_J^*\}$ in \eqref{eq_sec_rate}, we predict the instantaneous secrecy rate $\hat{C}_{S}^{(k^*)}(t+1),\ldots,\hat{C}_{S}^{(k^*)}(t+J)$. The prediction accuracy of the CSI and selected transmitter determines how close the estimated ESR $\hat{C}_{\mathrm{erg}}$ is from the actual ESR $C_{\mathrm{erg}}$.

\section{Numerical Results}
\label{section_results}

\begin{table}
\centering
\caption{Parameters for the dataset generation and training.}
\label{table_dataset}
\begin{tabular}{c|c}
\hline
Parameter  & \multicolumn{1}{c}{Value} \\ \hline
\hline
Carrier centre frequency $(f_c)$   &  2GHz\\
\hline
Sampling frequency $(f_s)$ &  1KHz\\
\hline
Channel fading   & Rayleigh\\
\hline
Number of paths   & 100\\
\hline
Doppler spectrum  & Clarke's model \cite{time_corr_chan_data_sim_twc_2005}\\

\hline
Historical CSI values $(T)$  & 10 \\
\hline
Predicted CSI values $(J)$  & 1 \\
\hline
Training data observations & 49990 \\
\hline
Test data observations & 149990 \\
\hline
Mini batch size  $(N_b)$ & 500\\
\hline
Training epochs  & 5 \\
\hline
Training iterations $(n\_iter)$   & 500 \\
\hline
Learning rate  & $\frac{0.005}{(1 + 0.005\times n_{iter})}  $ \\
\hline
Optimizer & Adam \\
\hline
Number of hidden units (LSTM) & 200 \\
\hline
Number of filters (CNN) & 50 \\
\hline
Filter size (CNN) & 6 \\
\hline
\end{tabular}
\vspace{-0.5cm}
\end{table}

 In this section, we evaluate the performance of the LSTM-based MTL architecture proposed in Fig. \ref{fig_ml_network_combined}. 
 We also compare its performance with that of a CNN-based MTL architecture.  
Both these MTL architectures are compared with the sequential task learning architecture shown in Fig. \ref{fig_ml_network_separate}. We use a typical personal computer with an Intel\textsuperscript{\textregistered} Core \textsuperscript{\texttrademark} i7-8700 CPU and 16 GB RAM for training purposes. We generate a time series CSI dataset 
using \eqref{eq_sd_chan_model} with specifications given  in Table \ref{table_dataset}. 
In all the figures, LSTM-J and CNN-J denote outputs from the MTL network architecture as proposed in Fig. \ref{fig_ml_network_combined}.  On the other hand, LSTM-S and CNN-S represent the outputs from the sequential network architectures shown in Fig. \ref{fig_ml_network_separate}. The CSI prediction error is shown in terms of NMSE defined in (\ref{eq_NMSE}) and transmitter selection accuracy
is shown as selection accuracy in the figures.

In Figs. \ref{fig_nmse_vs_weight} and \ref{fig_accuracy_vs_weight}, we show NMSE and the transmitter selection accuracy, respectively, for both the LSTM and CNN-based architectures by varying the weight $w$ in the MTL loss function. The figure tells how to choose $w$ for the proposed MTL problem. 
We observe that a higher weight assigned to $L_p$ decreases the CSI prediction error, however, if $w$ is more than 0.9, the transmitter selection accuracy drops. This suggests that the optimal weight for the problem is around $w=0.9$. However, the CSI prediction error and the transmitter selection accuracy remain relatively similar in the range of $0.1 \le w \le 0.9$. Hence, all the subsequent figures are plotted with $w =0.9$.

\begin{figure}
 \centering
  \begin{subfigure}{0.24\textwidth} 
  \centering 
  \includegraphics[width=\textwidth]{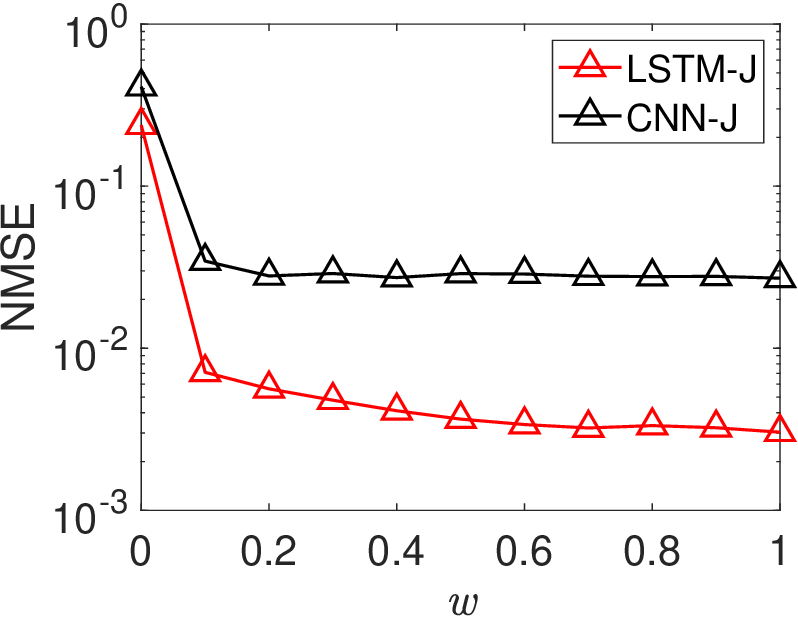}
 \caption{CSI prediction error.}
 \label{fig_nmse_vs_weight}
  \end{subfigure}
 \begin{subfigure}{0.24\textwidth}    \includegraphics[width=\textwidth]{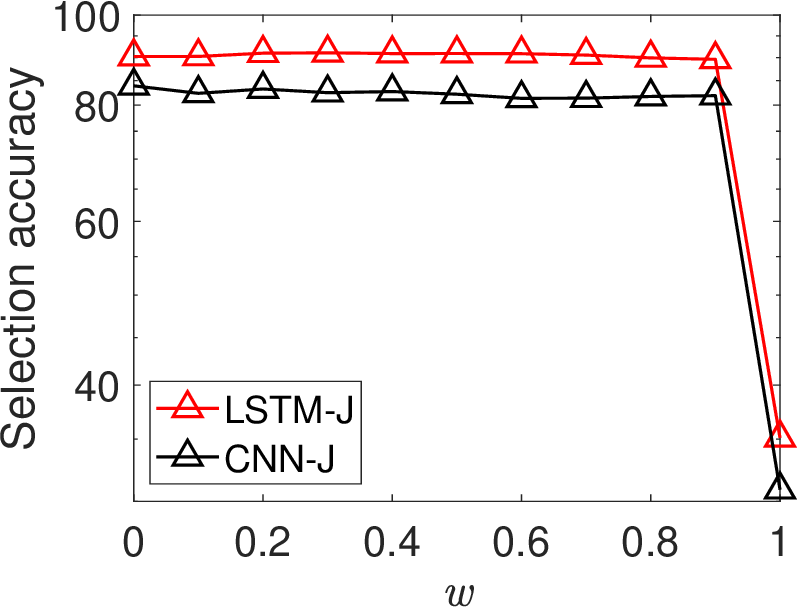}
 \centering
 \caption{Transmitter selection accuracy.}
 \label{fig_accuracy_vs_weight}
 \end{subfigure}
 \caption{Performance with $w$ when $\bar{\gamma}_{\mathrm{D}}=30$ dB, $\bar{\gamma}_\mathrm{E}=10$ dB, $K=3$, $N=2$, $v=10$ meters/second.}
 \label{fig_err_vs_weight}
 \vspace{-0.5cm}
 \end{figure}

In Figs. \ref{fig_nmse_vs_speed} and \ref{fig_accuracy_vs_speed}, we compare the CSI prediction error and the transmitter selection accuracy, respectively, for both the LSTM and CNN-based architecture by varying relative speed between transmitter and receiver $v$ from 5 to 50 meters/second. As seen in Fig. \ref{fig_nmse_vs_speed}, CSI prediction error increases with increasing $v$ for all the networks.
 With increasing $v$, the Doppler frequency increases, lowering the coherence time of the channel. 
 This weakens the correlation between CSI samples across multiple coherence intervals, making it difficult for both LSTM and CNN to learn the channel parameters.  We find that LSTM is better suited for CSI prediction in mobile scenarios as compared to CNN as it has significantly less CSI prediction error.  This is attributed to the LSTM's better ability to learn long-term dependencies in sequential data 
 than CNN \cite{jour_acm_cnn_lstm_comp_24}.

Due to reduced correlation between CSI samples with $v$ as described above, in Fig. \ref{fig_accuracy_vs_speed}, we see that the transmitter selection accuracy for both LSTM and CNN-based architectures also degrades with the increasing $v$.  
A better accuracy performance of LSTM is also observed as compared to CNN for a large mobility range. Finally, we observe that both MTL and sequential architectures provide similar CSI prediction errors. The transmitter selection accuracy of the MTL and sequential architectures are also comparable at higher $v$, however, the same for CNN-J is better than CNN-S at lower $v$. 

 \begin{figure}
 \centering
  \begin{subfigure}{0.24\textwidth}    \includegraphics[width=\textwidth]{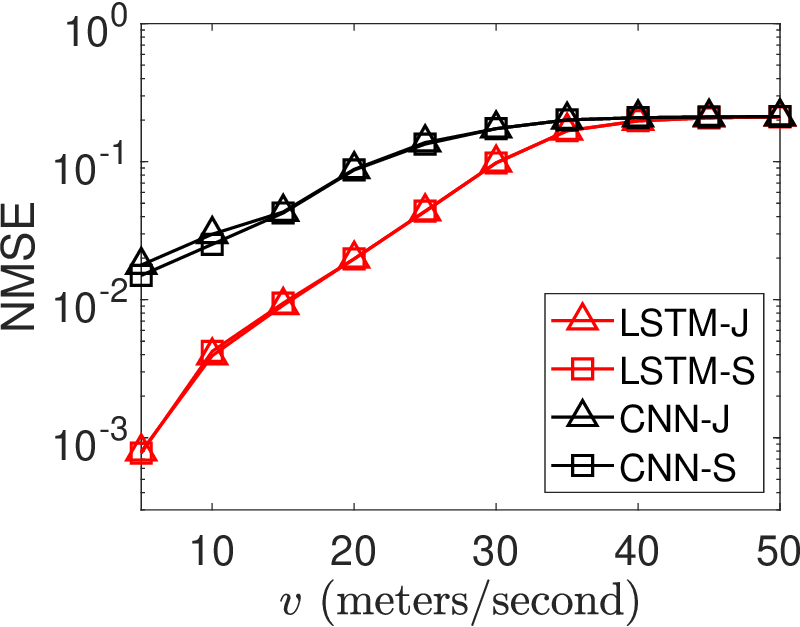}
 \caption{CSI prediction error.}
 \label{fig_nmse_vs_speed}
 \end{subfigure}
 \begin{subfigure}{0.24\textwidth}    \includegraphics[width=\textwidth]{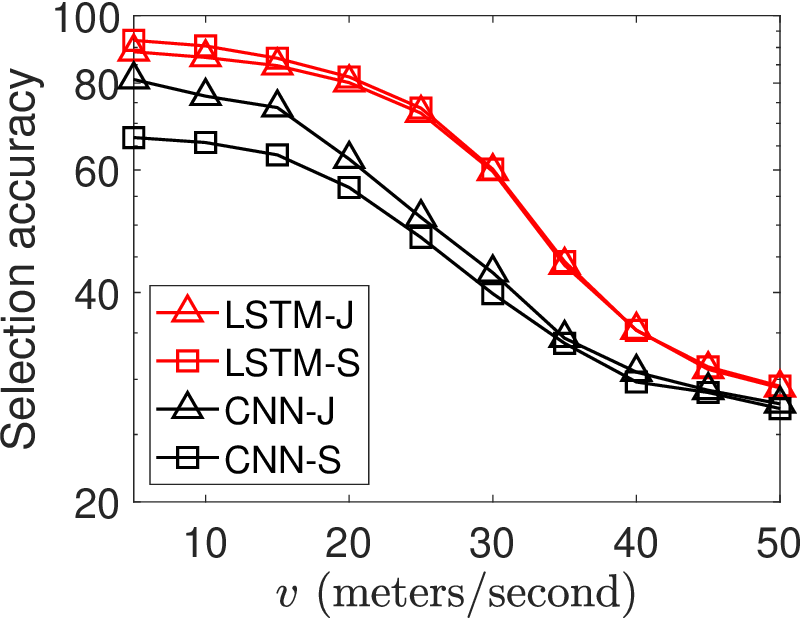}
 \caption{Transmitter selection accuracy.}
 \label{fig_accuracy_vs_speed}
 \end{subfigure}
 \caption{Performance with speed when $\bar{\gamma}_{\mathrm{D}}=30$ dB, $\bar{\gamma}_\mathrm{E}=10$ dB, $K=4$, $N=2$. }
 \label{fig_err_vs_speed}
 \vspace{-0.2cm}
 \end{figure}
 
In Figs. \ref{fig_nmse_vs_k} and \ref{fig_accuracy_vs_k}, the CSI prediction error and the transmitter selection accuracy, respectively, are plotted for the LSTM and CNN-based architecture by varying the number of transmitters $K$. 
In Fig. \ref{fig_nmse_vs_k}, we find that the increasing $K$ has a negligible effect on CSI prediction accuracy in all schemes. 
This observation is reasonable as transmitters are independent of each other. We also note that both MTL and
sequential architectures provide similar CSI prediction error with $K$. 
 
 
 In Fig. \ref{fig_accuracy_vs_k}, the selection accuracy degrades with increasing $K$. This is due to the growth in the learnable parameters with $K$ making it difficult to predict the selected transmitter accurately. 
 We observe that while CNN-J is always better than CNN-S, LSTM-J outperforms LSTM-S after a certain number of transmitters. 
 The advantage of MTL-based architectures (LSTM-J and CNN-J) in transmitter selection accuracy is that they can predict selected transmitters directly based on the historical CSI. The potential errors in sequential networks (LSTM-S and CNN-S) come from the fact that the individual tasks are performed sequentially. The accuracy of transmitter classification is thus dependent on the accuracy of CSI prediction.
 
\begin{figure}
 \centering
  \begin{subfigure}{0.24\textwidth}    \includegraphics[width=\textwidth]{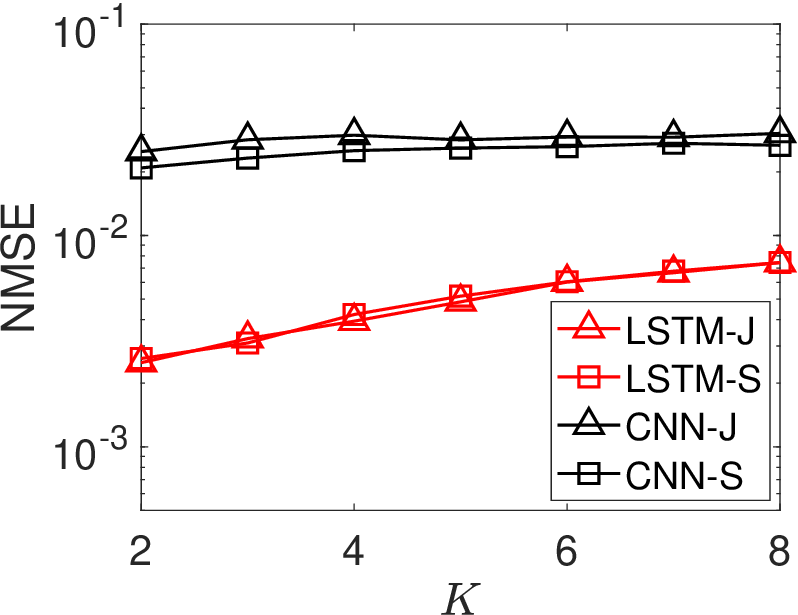}
 \caption{CSI prediction error.}
 \label{fig_nmse_vs_k}
 \end{subfigure}
   \begin{subfigure}{0.24\textwidth}    \includegraphics[width=\textwidth]{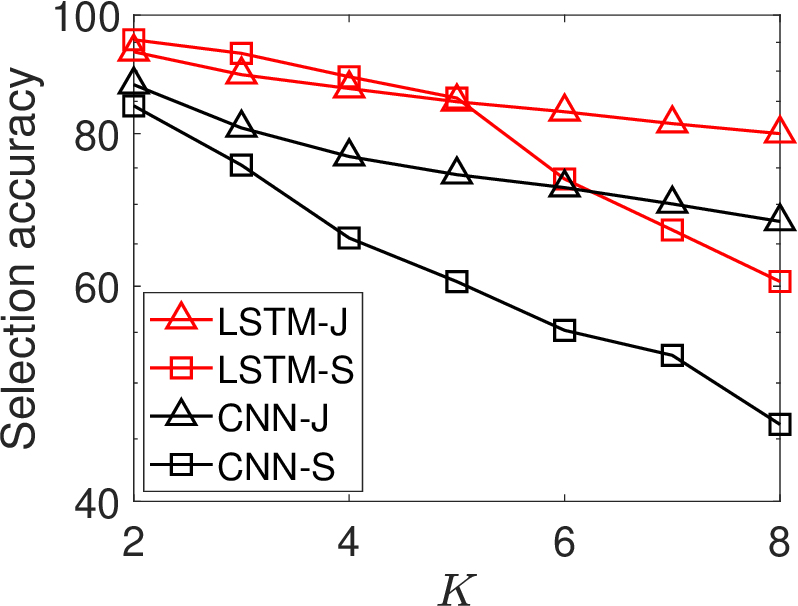}
 \caption{Transmitter selection accuracy.}
 \label{fig_accuracy_vs_k}
 \end{subfigure}
 \caption{Performance with the number of transmitters when $\bar{\gamma}_{\mathrm{D}}=30$ dB, $\bar{\gamma}_\mathrm{E}=10$ dB, $N=2$, and $v=10$ meters/second.}
 \label{fig_err_vs_k}
 \vspace{-0.5cm}
 \end{figure}
 
In Figs. \ref{fig_esr_vs_speed} and \ref{fig_esr_vs_k}, we depict the estimated ESR of proposed MTL-based networks (LSTM-J and CNN-J) versus $v$ and $K$, respectively. The estimated ESR is obtained by utilizing the estimated CSI and the predictive optimal transmitter produced by the MTL-based network in (\ref{eq_sec_rate}). The ESR of the MTL-based network is compared with that of the hypothetical perfect system with the perfectly estimated CSI and accurately selected optimal transmitter. 
It is clearly visible from both figures that the LSTM-based network is better than the CNN-based network in following the perfect system performance. 

 \begin{figure}
 \centering
  \begin{subfigure}{0.237\textwidth}  \includegraphics[width=\textwidth]{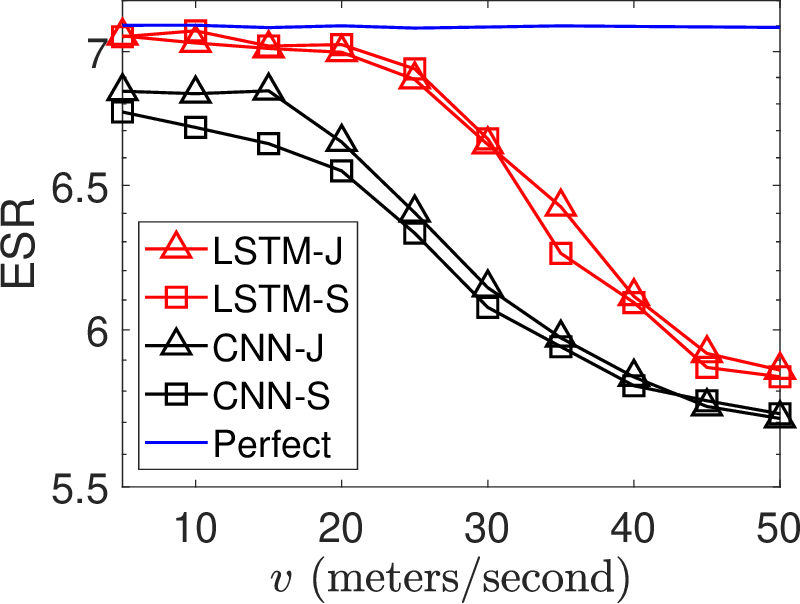}
 \caption{When $K=4$.}
 \label{fig_esr_vs_speed}
 \end{subfigure} 
  \begin{subfigure}{0.24\textwidth}  \includegraphics[width=\textwidth]{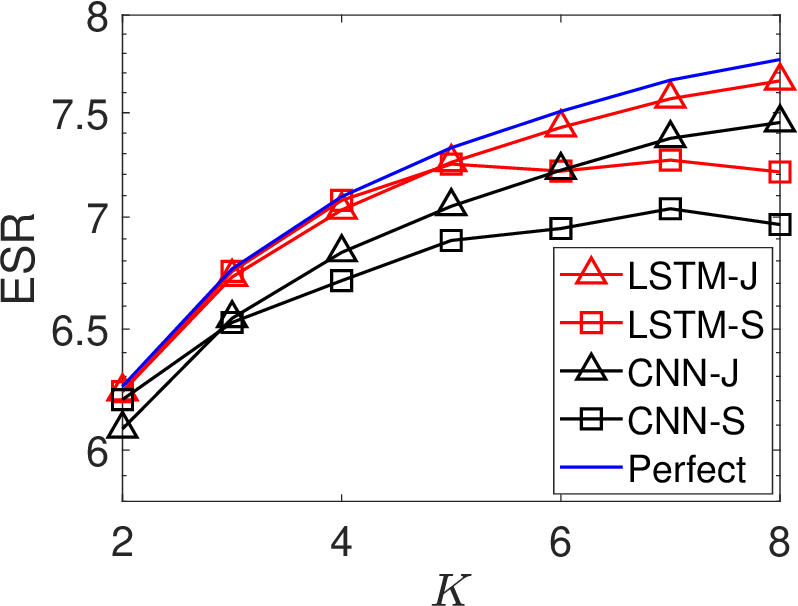}
 \caption{When $v=10$ meters/second.
 }
 \label{fig_esr_vs_k}
 \end{subfigure}
 \caption{ESR with $v$ and $K$ when $\bar{\gamma}_{\mathrm{D}}=30$ dB, $\bar{\gamma}_\mathrm{E}=10$ dB, $N=2$. }
  \vspace{-0.5cm}
\end{figure}
In Fig. \ref{fig_esr_vs_speed}, the ESR curves of both LSTM and CNN networks deviate significantly from the perfect system in the high-mobility region (high $v$). The ESR corresponding to MTL and sequential networks are close to each other for both LSTM and CNN. These observations are consistent with that in Fig. \ref{fig_nmse_vs_speed} and \ref{fig_accuracy_vs_speed}. 
When performance is considered with respect to $K$ in Fig. \ref{fig_esr_vs_k},
it is found that MTL-based network (CNN-J and LSTM-J) outperforms the sequential network beyond a certain $K$ in respective LSTM and CNN systems.  This observation is consistent with that in Fig. \ref{fig_nmse_vs_k}
and \ref{fig_accuracy_vs_k}.

A comparison of training and testing time required for the MTL and sequential model is shown in Table  \ref{table_train_test_time}. It is found that the training and testing time is reduced by 40.8 and 39.9 percent, respectively, in LSTM-J as compared to LSTM-S. A similar observation is also found in CNN-J versus CNN-S. Further, MTL-based architecture is trained only once, hence, it is also memory efficient as compared to the sequential task learning model which requires multiple trainings \cite{deep_learning_tut_ieee_com_survey_19}.

\begin{table}[]
\centering
\caption{Comparison in training and testing time (seconds).}
\label{table_train_test_time}
\begin{tabular}{c|c|c}
\hline
Network & Training time & Testing time \\ \hline \hline
LSTM-J  & 146.1         & 17.7         \\ \hline
LSTM-S  & 246.7         & 29.0         \\ \hline
CNN-J   & 102.5         & 10.1         \\ \hline
CNN-S   & 137.1         & 4.5          \\ \hline
\end{tabular}
\vspace{-.5cm}
\end{table}

\section{Conclusion}\label{section_conclusion}
In this paper, we propose a novel approach of joint CSI prediction and predictive transmitter selection tasks using historical CSI in mobile scenarios. This approach utilizes LSTM-based MTL architecture that exploits temporal correlation in CSI and the inherent relationship between tasks. Compared to the CNN architecture, LSTM provides better performance. Compared to the sequential task learning model MTL architecture provides superior predicted secrecy performance for large variations in the number of transmitters and mobile node speeds. It also delivers better computational and memory efficiency. The LSTM-based MTL architecture is found to reduce the computational time by around 40 percent as compared to its sequential task learning architecture. 

\bibliographystyle{IEEEtran}
\bibliography{IEEEabrv, ref}
}
\end{document}